\documentclass[a4paper, 11pt]{article}
\pdfoutput=1

\usepackage{amsmath,amssymb,graphicx,multirow,xspace}
\usepackage[colorlinks=true,urlcolor=blue,anchorcolor=blue,citecolor=blue,filecolor=blue,linkcolor=blue,menucolor=blue,pagecolor=blue,linktocpage=true]{hyperref}
\usepackage[compress,numbers]{natbib}
\usepackage{braket}
\usepackage{slashed}
\usepackage{physics}
\usepackage{bm}
\usepackage{indentfirst}
\usepackage{appendix}
\usepackage{verbatim}
\usepackage{cancel}
\usepackage{graphicx}
\usepackage{subcaption}

\usepackage[T1]{fontenc}
\long\def\symbolfootnote[#1]#2{\begingroup%
  \def\thefootnote{\fnsymbol{footnote}}\footnote[#1]{#2}\endgroup}

\newcommand{\gev}{\mathrm{GeV}}

\newcommand{\hvev}[1]{ \left\langle {#1} \right\rangle }

\input prepictex
\input pictex
\input postpictex
\newdimen\tdim
\tdim=\unitlength

\begin{document}

\begin{titlepage}

   \begin{flushright}
     \small{CERN-TH-2022-107}\
   \end{flushright}

  \vspace{0.5cm}
  \begin{center}
    \Large\bf
    First-Order Electroweak Phase Transition and\\Baryogenesis from a Naturally Light Singlet Scalar
  \end{center}

  \vspace{0.2cm}
  \begin{center}
    {
      Keisuke Harigaya$^{1}$\symbolfootnote[1]{keisuke.harigaya@cern.ch}
      and Isaac R. Wang$^2$\symbolfootnote[2]{isaac.wang@rutgers.edu}
    }\\
    \vspace{0.6cm}
    \textit{$\,^1$ Theoretical Physics Department,
      CERN,\\
      1211 Geneva 23, Switzerland\\
      \vspace{0.5cm}
      $\,^2$New High Energy Theory Center,
      Department of Physics and Astronomy,\\
      Rutgers University, Piscataway, NJ 08854, USA}\\
  \end{center}

  \vspace{0.4cm}

  \begin{abstract}
    We investigate a minimal singlet-scalar extension to the Standard Model that achieves a strong first-order electroweak phase transition. The singlet can be naturally light because of an approximate shift symmetry and no extra hierarchy problem beyond that of the Standard Model Higgs is introduced. We discuss the two-field dynamics of the phase transition in detail and find that the gravitational-wave signal is too weak to be detected by near-future observations. We also discuss the meta-stability of the zero-temperature scalar potential. Despite the apparent instability just above the electroweak scale, we show that the lifetime of the electroweak vacuum is much longer than the age of the universe and hence the setup does not require UV completion near the electroweak scale. The baryon asymmetry of the universe may be explained by local electroweak baryogenesis arising from a coupling between the singlet and weak gauge boson. The predicted electron electric dipole moment is much below the current bound. The viable parameter space can be probed by the observations of rare Kaon decay and the cosmic microwave background.
  \end{abstract}

\end{titlepage}

\vspace{0.2cm}
\noindent

\tableofcontents
\newpage

\section{Introduction}
\label{sec:introduction}

The observed matter-antimatter asymmetry remains an unsolved puzzle for particle physics and cosmology.
Among various scenarios, electroweak baryogenesis (EWBG)~\cite{Kuzmin:1985mm} utilizes a baryon number-violating process that already exists in the Standard Model (SM), called the electroweak sphaleron process~\cite{Manton:1983nd,Klinkhamer:1984di}, to generate baryon asymmetry in the early universe and is particularly interesting.
The electroweak phase transition (EWPT), if first-order, provides a departure from thermal equilibrium. With CP violation, all of the necessary conditions to generate baryon asymmetry~\cite{Sakharov:1967dj} can be satisfied.
However, the EWPT in the SM has been proved to be a smooth crossover~\cite{Jansen:1995yg,Kajantie:1995kf,Rummukainen:1996sx,Kajantie:1996mn,Gurtler:1997hr,Csikor:1998eu,Laine:1998vn,Laine:1998qk,Rummukainen:1998as,Fodor:1999at}, which excludes EWBG within the SM. Furthermore, the smallness of the CP violation in the SM suppresses the amount of baryon asymmetry produced by EWBG in the SM even if the EWPT is of first order~\cite{Huet:1994jb,Gavela:1993ts,Gavela:1994ds,Gavela:1994dt,Gavela:1994yf}. For reviews of EWBG or other baryogenesis scenarios, see~\cite{Cohen:1993nk,Dine:2003ax,Cline:2006ts,Morrissey:2012db,Elor:2022hpa,Asadi:2022njl}. For reviews of EWPT, see~\cite{Morrissey:2012db,Quiros:1999jp,Postma:2020toi,Croon:2020cgk}.

New physics is thus required for successful EWBG. A strong first-order EWPT (SFOPT) may be achieved by extending the SM with an extra singlet scalar that generates tree-level trilinear terms~\cite{Pietroni:1992in,Choi:1993cv,Ham:2004cf,Noble:2007kk,Ahriche:2007jp,Profumo:2007wc,Barger:2007im,Barger2008,Das:2009ue,Ashoorioon:2009nf,Barger:2011vm,Espinosa:2011ax,Chung:2012vg,Profumo:2014opa,Kotwal:2016tex,Tenkanen:2016idg,Chen:2017qcz,Carena:2019une,Friedlander:2020tnq}. (A trilinear term induced by the thermal-loop of singlets can also enable SFOPT. See~\cite{Anderson:1991zb,Espinosa:1993bs} for the early literature on this scenario.) These singlet extensions introduce a scalar field that is singlet under the SM gauge groups but interacts with the SM Higgs field, depending on the symmetry that is assumed. Mixing with the SM Higgs field generates tree-level trilinear interaction
that contributes to the thermal barrier of the Higgs potential at finite temperature, which makes the EWPT of first order.
Additional sources of CP violation that enable EWBG have been considered in the singlet extensions~\cite{McDonald:1993ey,Espinosa:2011eu,Cline:2012hg,Cheung:2013dca,Jiang:2015cwa,Vaskonen:2016yiu,Cline:2017qpe,Grzadkowski:2018nbc,Carena:2019xrr,Chao:2019smr,Xie:2020bkl,Cline:2021iff}. Typically, these models with singlets
predict signals that can be probed by collider experiments, electric dipole moments (EDMs), and gravitational waves (GWs).

In order for the singlet to affect the EWPT, the singlet mass must be around or below the electroweak scale. This generically introduces an extra hierarchy problem in addition to that of the Higgs mass parameter; why is the singlet mass much smaller than the fundamental scale of the theory, such as the Planck scale?
The lightness of the singlet may be explained by the same dynamics that explains the smallness of the electroweak scale, such as supersymmetry and composite Higgs. Although one of these may be the right way to understand the smallness of the singlet mass, given that we have not yet found the trace of these theories at collider experiments, it will be useful to pursue another possibility to control the singlet mass.

The singlet mass may be controlled by a shift symmetry $S\rightarrow S+\delta S$ and soft breaking of it.
The leading shift symmetry-breaking terms are the mass term $\mu_S^2 S^2$ and a trilinear coupling
$A S |H|^2$.
The trilinear coupling generates quantum corrections to the mass
\begin{align}
  \label{eq:trilinear correction}
  \delta \mu_S^2 \simeq \frac{A^2}{16 \pi^2}{\rm ln}(\frac{\Lambda_{H}}{v}),
\end{align}
where $\Lambda_H$ is the cut-off scale of the Higgs loop.
This correction is proportional to $A^2$,
and thus the extra hierarchy problem is avoided as long as $A \lesssim \mu_S$. The potential with such an interaction is written as
\begin{align}
  \label{eq:model}
    V_0 = - \mu_H^2|H|^2 + \lambda |H|^4 + \frac{1}{2}\mu_S^2S^2 - A S (|H|^2-v^2).
\end{align}
Here $v = \mu_H/\sqrt{2\lambda} \simeq 174~\gev$ is the observed Higgs vacuum expectation value (vev). We shifted $S$ to chose the tadpole term $ASv^2$ and set $\hvev{S} = 0$ at $T=0$. All parameters are real and positive. There are only two extra free parameters beyond the SM.

The model in Eq~\eqref{eq:model} was first introduced in~\cite{Das:2009ue} and further reviewed in~\cite{Espinosa:2011ax}, motivated by the minimality. Here, we point out that this model has an extra motivation from the naturalness perspective.
Ref.~\cite{Espinosa:2011eu} identifies a singlet with a Nambu-Goldstone boson in a non-minimal composite Higgs model~\cite{Gripaios:2009pe}, so the lightness of the singlet is also understood by a shift symmetry. In their setup, self couplings of the singlet and the quartic interaction between the singlet and the Higgs are introduced. In our setup, the singlet is not embedded into composite Higgs scenarios. Indeed, as we will see in Sec.~\ref{sec:ewbg and edm},  successful EWBG points towards the decay constant of the singlet far above the EW scale.

We analyse the phase-transition dynamics of the model in detail.
The previous work is based on a 1-dimensional approximated analysis by integrating out the light scalar, where an observable amount of GWs is predicted. In this work, we show that including the 2-field dynamics is necessary in order to obtain a correct understanding of the finite temperature phase transition and vacuum tunneling. Including 2-field dynamics in phase transition leads to suppressed gravitational-wave signals and a different preferred parameter space, while that in vacuum tunneling protects the vacuum metastability, despite the apparent existence of a deeper vacuum near the electroweak vacuum.

EWBG may be achieved by the coupling of the singlet with the $W$ boson, as in the local EWBG scenario with a Higgs-$W$ boson coupling~\cite{Dine:1990fj}. Typically, singlet-scalar extensions of the SM aiming at EWBG lead to  2-loop corrections to the EDM of the SM fermions and are severely constrained by the electron EDM~\cite{Espinosa:2011eu,Chao:2019smr}. In our setup, we find that the EDM is strongly suppressed since the required magnitude of the trilinear coupling is small for a small singlet mass.

Successful EWBG puts an upper bound on the cut-off scale of the singlet-W boson coupling. The singlet must be UV-completed at this scale. There may be corrections to the singlet mass at the UV scale beyond that can be captured in the low energy effective theory in Eq.~\eqref{eq:model}. We present a toy example where quantum corrections to the singlet mass at the UV scale are suppressed without the aid of the possible compositeness of the SM Higgs or supersymmetry. In this theory, no significant modification of the SM sector is required for the lightness of the singlet.

We also investigate other experimental probes. 
The viable parameter region predicts the singlet scalar mass around 10 MeV. This parameter region has too small mixing between the singlet and the Higgs to be probed by future colliders.
Instead, the parameter region can be probed by the observations of 
rare Kaon decay and the cosmic microwave background (CMB). This is quite different from other EWBG models with a singlet that can be probed by future colliders, EDMs, and GWs.

This paper is organized as follows. In Sec~\ref{sec:model}, we discuss the basic settings of the model. In Sec~\ref{sec:phase transition}, we discuss the thermal phase transition in detail, including the 2-field dynamics. In Sec~\ref{sec:signal}, we discuss experimental and observational probes of the model. In Sec~\ref{sec:ewbg and edm}, we discuss how EWBG can be achieved in this model and the implications to the UV completion of the model. In Sec~\ref{sec:metastability}, the vacuum stability of the Higgs potential is discussed. Sec.~\ref{sec:discussion} is devoted to conclusions and discussion.

\section{Model}
\label{sec:model}
The scalar potential of the Higgs $H$ and a new light singlet scalar $S$ is given in Eq~\eqref{eq:model}.
The Higgs field is decomposed as
\begin{align}
  \label{eq:Higgs component}
  H = \frac{1}{\sqrt{2}}\begin{pmatrix}\chi_1 + i \chi_2\\h+i\chi_3\end{pmatrix},
\end{align}
where $\chi_{1,2,3}$ are would-be Nambu-Goldstone modes, $h$ is the physical Higgs field that obtains a non-zero vev $\hvev{h} = \sqrt{2}v$ at $T=0$. The tree-level potential of $h$ and $S$ is
\begin{align}
  \label{eq:tree-level potential}
  V_0(h,S) = - \frac{1}{2}\mu_H^2 h^2 + \frac{1}{4}\lambda h^4 + \frac{1}{2}\mu_S^2 S^2 - \frac{1}{2}A S (h^2-2v^2).
\end{align}
The free parameters of the potential are $\{\lambda, \mu_H, \mu_S, A\}$. We may translate them to the following four parameters; the Higgs mass $m_H = 125~\gev$, light scalar mass $m_S$, Higgs vev $v=174~\gev$, and mixing angle $\theta$ between the doublet and singlet.
See Appendix~\ref{sec:parametrization} for the correspondence of the parameters.
The vacuum stability requires that
\begin{align}
  \label{eq:denominator positive}
  2\lambda \mu_S^2 - A^2 > 0.
\end{align}

The possibility of the first order EWPT in this setup can be intuitively understood in the following way. Along the trajectory that minimizes the potential energy for a given $H$,  where $\partial V/ \partial S=0$,
\begin{align}
    V(h,\frac{A h^2}{2 \mu_S^2}) \simeq - \frac{1}{2} \mu_H^2|H|^2 + \frac{1}{4}(\lambda - \frac{A^2}{2 \mu_S^2})h^4.
\end{align}
If $A^2 \sim 2 \lambda \mu_S^2$, we may make the effective quartic coupling to be $O(0.01)$. Here it is crucial that the singlet mass is below the Higgs mass; otherwise we may indeed integrate out the singlet and the small quartic coupling is no longer an effective one but is the one to determine the physical Higgs mass.
With the trilinear thermal potential given by the weak gauge boson loop, a first order EWPT may be achieved, as in the SM with a smaller Higgs mass. More careful treatment requires solving two-field dynamics at finite temperature, which will be discussed in Sec~\ref{sec:phase transition}, but the smallness of the tree-level effective quartic coupling provides an intuitive understanding.

Assuming $\mu_S, A \ll v$, the physical mass eigenvalues are
\begin{align}
  \label{eq:physical scalar masses}
    m_+^2 \equiv m_h^2 \simeq & 4 \lambda v^2 (1+ \frac{A^2}{8 \lambda^2 v^2}), \nonumber \\
    m_-^2 \equiv m_S^2 \simeq & \mu_S^2  - \frac{A^2}{2 \lambda},
\end{align}
where $m_h = 125~\gev$ is the observed SM Higgs mass while $m_S$ is the mass of the light scalar. Eq.~\eqref{eq:denominator positive} ensures that $m_-^2$ is positive, which provides another way to understand this constraint.

Since $\lambda \simeq 0.1$, the effective quartic coupling of $O(0.01)$ requires tuning of $10 \%$.
Here we define the tuning by
\begin{align}
\label{eq:tuning}
    \frac{\lambda- A^2/2\mu_S^2 }{\lambda} = 1 - \frac{A^2}{2 \lambda \mu_S^2} = \frac{m_S^2}{\mu_S^2}.
\end{align}

\section{Thermal correction and strong first order electroweak phase transition}
\label{sec:phase transition}
In this section, we analyse the phase-transition dynamics of the singlet extension. We show that the duration of the phase transition becomes short for $m_S \ll m_h$. This implies that in the viable parameter region discussed in Sec.~\ref{sec:signal}, the magnitude of gravitational waves produced during the first order phase transition becomes weak.

\subsection{Thermal potential}

We discuss the thermal corrections to the potential of the Higgs and singlet following~\cite{Das:2009ue}.
At high temperatures, the effective potential receives thermal corrections. At one-loop level, the effective potential can be written under the high temperature expansion as
\begin{align}
  \label{eq:highT potential form}
  V = D(T^2 - T_0^2)h^2 - E_{\rm SM} T h^3 + \frac{1}{4}\lambda h^4 + \frac{1}{2}\mu_S^2S^2 -\frac{1}{2} A (h^2 + c_s T^2-2v^2)S - E(h,S)T,
\end{align}
where the coefficients are given by
\begin{align}
  \label{eq:highT coefficients}
  D &= \frac{1}{16}\left (3g^2 + g^{\prime 2} + 4y_t^2 - 4 \lambda \right ),
  ~~T_0^2 = \frac{\mu_H^2}{2D},\nonumber \\
  E_{\rm SM} &= \frac{1}{48\pi}\left (2g^3+(g^2 + g^{\prime 2})^{3/2}  \right ) = \frac{1}{12\sqrt{2}\pi v^3}\left (2m_W^3 + m_Z^3  \right ), 
  ~~c_s = \frac{1}{3}, \nonumber \\
  E(h,S) &= \frac{1}{12\pi} \left (3(-\mu_H^2 -  A S +  \lambda h^2)^{3/2} + \left(m_h^2(h,S)\right)^{3/2} + \left (m_S^2(h,S)\right)^{3/2}  \right ).
\end{align}
If we neglect the scalar contribution to all of these coefficients except for the trilinear term $c_s$, the potential has the form of $V_{\rm SM} - AS(h^2 -2v^2+ c_sT^2)/2$. In this case, $D$ becomes $D_{\rm SM} = (2m_W^2 + m_Z^2 + 2m_t^2)/(16v^2)$ and $E(h,S)T$ term is dropped. This is indeed a good approximation since the couplings $\lambda$ and $A$ are small. Note that we apply a factor of $2/3$ to the trilinear term $E_{\rm SM}$ to screen out the longitudinal mode. We may instead include the longitudinal mode while performing the resummation~\cite{Quiros:1999jp,Dine:1992wr}.
We find that the two methods show great agreement with each other.

For any temperature $T$, the vev of $S$ for a given $h$ is given by
\begin{align}
  \label{eq:general S vev}
  \hvev{S}(h,T) = \frac{A}{2\mu_S^2}(h^2 + c_s T^2-2v^2),
\end{align}
which is always $T$-dependent. This means that the thermal history of the vevs during the cooling of the universe is a one-step transition with two-field dynamics involved.
The critical temperature and the widely-used definition of the strength of the phase transition are thus given by
\begin{align}
  \label{eq:highT Tc}
  T_c^2 = \frac{ \lambda^{\prime} D^{\prime}}{ \lambda^{\prime} D^{\prime} - E_{\rm SM}^2} T_0^{\prime 2},~~~
  \frac{v_c}{T_c} = \frac{2E_{\rm SM}}{\lambda^{\prime}},
\end{align}
where
\begin{align}
  \label{eq:effective high T coefficient}
  \lambda^{\prime} \equiv \lambda - \frac{A^2}{2\mu_S^2},~~
  D^{\prime} \equiv D_{\rm SM} - \frac{A^2}{4\mu_S^2}c_s,~~
  T_0^{\prime 2} \equiv \frac{\mu_H^2\mu_S^2 - A^{2} v^2}{2 D^{\prime}\mu_S^2}.
\end{align}
We note that the definition of the phase-transition strength in Eq.~\eqref{eq:highT Tc} may not be accurate because of the difference between the critical temperature and the nucleation temperature that becomes larger for smaller $m_S$. We discuss it shortly in Sec.~\ref{sec:Tn and Tc}. This expression, however, provides a simple and intuitive understanding of the phase-transition strength.
The impact of the extra scalar $S$ is readily understood from Eq.~\eqref{eq:highT Tc}; the small effective quartic, $\lambda'$, directly contributes to stronger phase transition. If we include the extra contribution $E(h,T)$, the phase transition strength is further enhanced, but only slightly.

\subsection{Effect of Coleman-Weinberg correction to the phase transition}
\label{sec:drop CW}

The effective potential also receives zero-temperature quantum corrections known as the Coleman-Weinberg (CW) correction, written as
\begin{align}
  \label{eq:CW potential}
  V_{\rm CW} = \frac{1}{64\pi^2}\left( \sum_B m_B^4(h)\left[ \log \left( \frac{m_B^2(h)}{Q^2} \right)-c_B \right] - \sum_Fm_F^4(h) \left[ \log \left( \frac{m_F^2(h)}{Q^2} \right)-c_F \right] \right),
\end{align}
where $B$ ($F$) are bosonic (fermionic) degrees of freedom.
The constants $c_B$ and $c_F$ are scheme-dependent, but the dependence on them disappears after choosing the tree-level parameters while fixing the physical parameters.
We include gauge bosons but neglect scalar-boson contributions due to the smallness of the latter couplings.
The only fermion that matters here is the top quark. All of those masses do not depend on $S$. In summary, the particles and their masses that enter the above equations are
\begin{align}
  \label{eq:particle list}
  m_{W}^2(h) &= \frac{1}{4}g^2h^2,\ n_{W} = 6,\nonumber \\
  m_{Z}^2(h) &= \frac{1}{4}(g^2+g^{\prime 2})h^2,\ n_{Z} = 3,\nonumber \\
  m_t^2(h) &= \frac{1}{2}y_t^2h^2,\ n_t= 12.
\end{align}

In general, this correction should be included in full-numerical computation. However, the large top Yukawa coupling makes the 1-loop corrected potential severely deviate from the tree-level one, and thus causes an irregularly large impact on the phase-transition strength. The situation is further exacerbated for a smaller quartic coupling. This is similar to what happens in the SM.
In the SM, lattice simulation~\cite{Kajantie:1995kf,Kajantie:1996mn,Gurtler:1997hr,Csikor:1998eu,Rummukainen:1998as} has shown that the strong first order phase transition can be achieved for $m_h \lesssim 43~\gev$ for a top quark mass $m_t \simeq 173~\gev$. The high-temperature expanded perturbation theory without the 1-loop CW correction shows good agreement with this conclusion~\cite{Quiros:1999jp,Dine:1992wr}. The computation with the 1-loop CW correction, however,
predicts first order phase transition
only with top quark mass in the range $120\sim 150~\gev$, where the correction from the top quark is small and 
the CW correction causes disagreement of the perturbative calculation with the lattice simulation for a top quark mass larger than $150~\gev$~\cite{Dine:1992wr}.
This suggests that the perturbative calculation should be performed to higher order loops~\cite{Kajantie:1995kf} with more techniques involved, e.g., dimension reduction (for recent progress and reviews, see~\cite{Croon:2020cgk,Schicho:2021gca,Schicho:2022wty,Niemi:2021qvp}).

Instead of performing higher order computations, we simply use the high-temperature expansion without the 1-loop CW correction to analyze the finite temperature phase-transition dynamics in the minimal singlet extension.
Indeed, the minimal singlet extension mimics the SM with a small quartic coupling, so we expect that the method that reproduces the precise lattice prediction in the SM also gives a correct answer in our setup.  
The CW correction also impacts the metastability of the electroweak vacuum at the zero-temperature, which will be discussed in detail in Sec~\ref{sec:metastability}.

\subsection{Two-dimensional dynamics of bubble nucleation}
\label{sec:2dpt}

First-order phase transition is induced via bubble nucleation. Starting from the critical temperature $T_c$, bubble nucleation has a non-zero rate that depends on the action. As the universe cools down, this action becomes smaller and the tunneling rate increases. Bubble nucleation happens once the nucleation rate matches the Hubble rate at $T=T_n$, called the nucleation temperature, satisfying
\begin{align}
  \label{eq:nucleation action}
 \left. \frac{S_3}{T}\right\vert_{T_n} \simeq 140,
\end{align}
where $S_3$ is the $O(3)$ symmetric bounce action of the bubble nucleation process, which has the largest contribution to the nucleation rate.

In general, the phase-transition dynamics should be solved in a multi-dimensional field space. The transition dominantly occurs along the path in the field configuration space that minimizes the bounce action $S_{3}$. We find that the solution in the 2d space for this model closely follows the ``valley'' where the $S$ field value is minimized at finite temperature, i.e., Eq.~\eqref{eq:general S vev}. 

One may then guess that performing computation in the 1d space of $H$ by integrating out the light scalar $S$ can simplify the computation without making a significant error.
Indeed, in Ref.~\cite{Das:2009ue}, the authors performed the analysis in the 1d approximation. After fixing the singlet field value with Eq.~\eqref{eq:general S vev}, the effective potential becomes
\begin{align}
  \label{eq:1d potential}
  V_{\rm 1d}(h) = D^{\prime}(T^2 - T_0^{\prime 2})h^2 - E_{\rm SM}T h^3 + \frac{\lambda^{\prime}}{4}h^4,
\end{align}
where $D^{\prime}$, $T_0^{\prime}$ and $\lambda^{\prime}$ are given in Eq~\eqref{eq:effective high T coefficient}.
The approximated action $S_{3, \rm 1d}$ is 
\begin{align}
  \label{eq:1d action}
  S_{3, \rm 1d} = 4\pi \int r^2 \left( \frac{1}{2}\left(\frac{dh(r)}{dr}\right)^2 + V(h) \right)dr,
\end{align}
where $h(r)$ is the Higgs field on the bubble. Note that only the gradient energy of the $h$ field contributes to the action.
We find that this approximation gives a result that agrees with the full 2D computation for $m_S \gg 1$ GeV, but fails for smaller $m_S$.
The lightness of the extra scalar leads to a large change of the field value of $S$ during the transition, and thus leads to a large gradient-energy contribution to the bounce action, which delays the bubble nucleation. This effect is neglected in the 1d approximation.

\begin{figure}[t]
  \centering
  \begin{minipage}[t]{0.495\linewidth}
    \hspace{0.5cm}
    \includegraphics[width=3.2in]{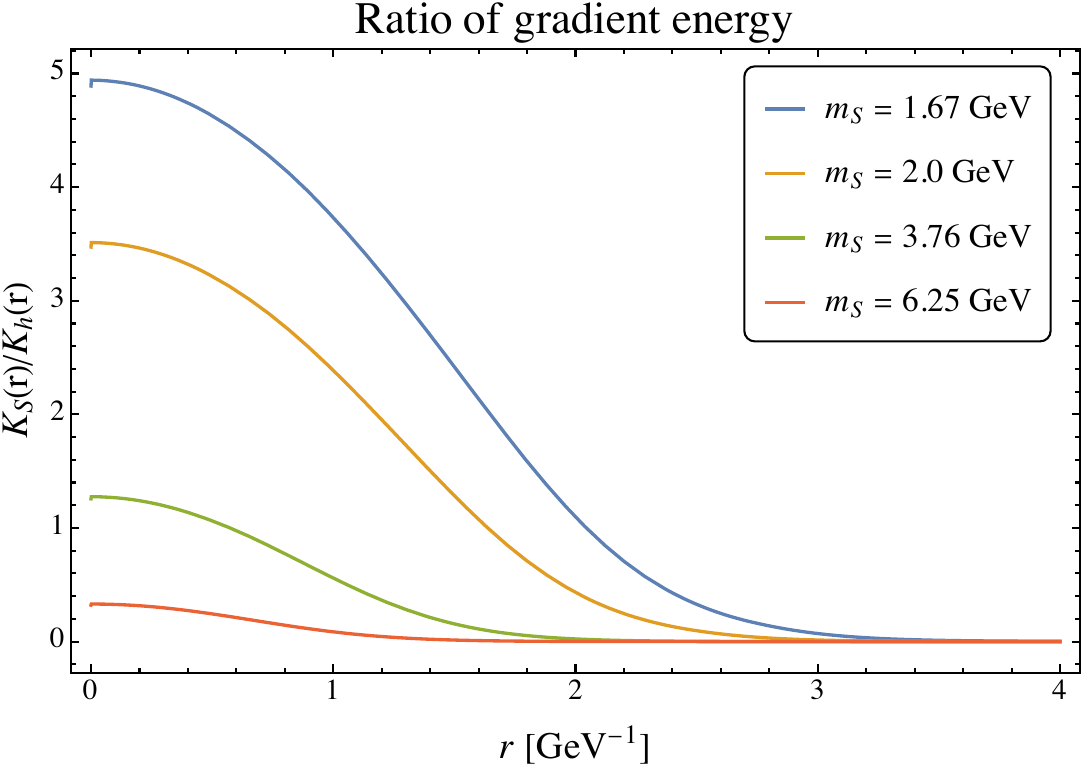}
  \end{minipage}
  \begin{minipage}[t]{0.495\linewidth}
    \hspace{0.5cm}
    \includegraphics[width=3.2in]{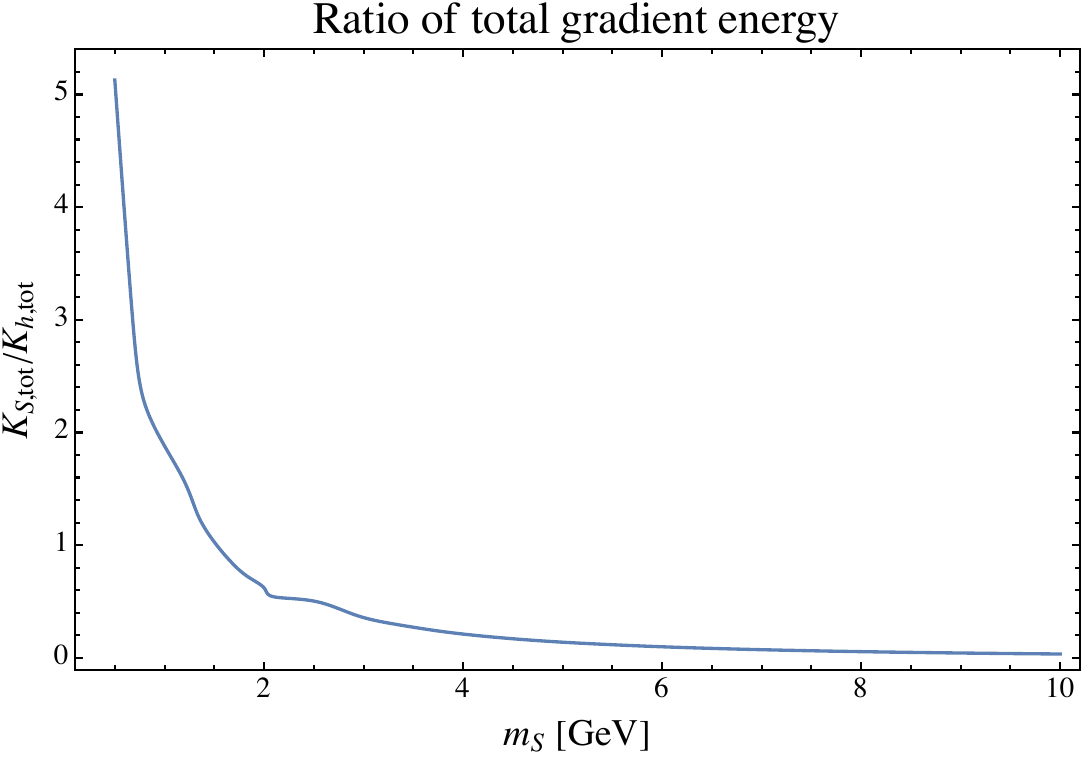}
  \end{minipage}
  \caption{Ratio of the contribution to the bounce action from the gradient terms $K_S/K_h$ in the full 2d computation at the nucleation temperature of the 1d approximation $T_{\rm n,1d}$. Left: The ratio on the bubble profile for several values of $m_S$. Right: Gradient energy integrated over the bubble profile and scanned over different masses. The fine-tuning is fixed at $5\%$ to determine the mixing angle.}
  \label{fig:Kratio}
\end{figure}

To explicitly see this, let us check the gradient-energy contribution to the bounce action,
\begin{align}
  \label{eq:kinetic energy definition}
  4\pi \int \frac{1}{2}r^2 \left( \left(\frac{dh(r)}{d r}\right)^2 + \left(\frac{dS(r)}{dr}\right)^2 \right) dr \equiv 4\pi \int \frac{1}{2}r^2 (K_h(r) + K_S(r)) dr \equiv K_{h,\rm tot} + K_{S, \rm tot},
\end{align}
where $h(r)$ and $S(r)$ are the bubble profiles in the 3d Euclidean space. Along the path, i.e., Eq.~\eqref{eq:general S vev}, the ratio of the gradient terms of $S$ and $h$ is given by
\begin{align}
  \label{eq:kinetic energy ratio}
  \frac{K_S(r)}{K_h(r)} =\frac{A^2}{\mu_S^4}h(r)^2.
\end{align}
One can see that smaller $\mu_S$ leads to larger contribution from the gradient energy of $S$, which invalidates the 1d approximation.
This extra contribution to the action leads to delayed bubble nucleation, i.e., a smaller $T_n$, compared to the 1d approximation.
For a better comparison, we first obtain the nucleation temperature in the 1d approximation, $T_{\rm n, 1d}$, and then solve the 2d tunneling at this temperature to see how the action is changed.
All computations of the phase transition and thermal tunneling at finite temperature are performed with the help of \verb|CosmoTransitions|~\cite{Wainwright:2011kj}, with some modification to analyse our setup.
In the left-panel of Fig~\ref{fig:Kratio}, we show the ratio $K_S(r)/K_h(r)$ from the solution of the 2d tunneling at $T_{\rm n, 1d}$ for different scalar masses with sin$\theta$ fixed so that the fine-tuning in Eq.~\eqref{eq:tuning} is $5\%$. The analytical expression of this ratio $A^2h^2/\mu_S^4$ agrees with the result of the numerical solution. In the right panel of Fig~\ref{fig:Kratio}, we show the ratio $K_{S,\rm tot}/K_{h,\rm tot}$ as a function of scalar masses with fine-tuning still fixed to be $5\%$. As the scalar mass $m_S$ goes to zero, the ratio $K_{S,\rm tot}/K_{h,\rm tot}$ effectively diverges.

\begin{figure}[t]
  \centering
  \includegraphics[width=0.6\linewidth]{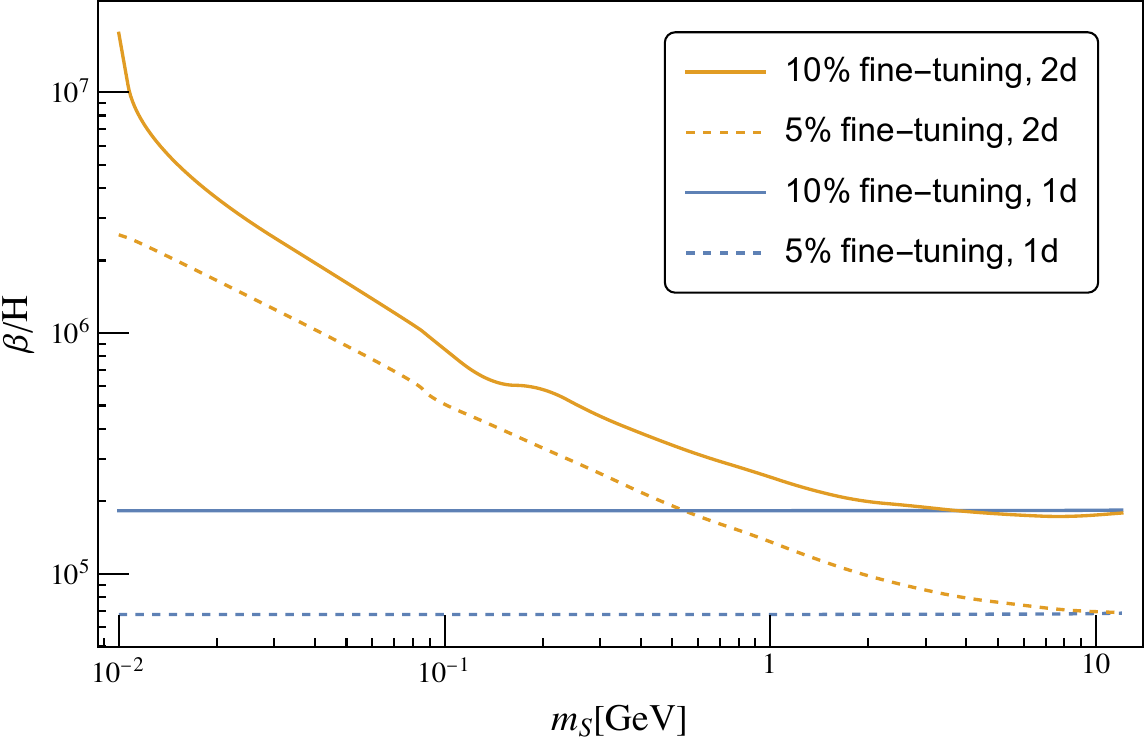}
  \caption{The inverse time duration parameter $\beta/H$ as a function of the scalar mass $m_S$ in the full 2d computation and 1d approximation. A comparison is made for fixed fine-tuning of $5\%$ and $10\%$.}
  \label{fig:betaH compare fix tuning}
\end{figure}

Another important quantity to describe the electroweak phase transition is the inverse time duration, $\beta$ parameter, which is usually normalized by the Hubble expansion rate $H$ and is given by
\begin{align}
  \label{eq:betaH def}
  \frac{\beta}{H} = T_n \left.\frac{d S_3/T}{d T}\right\vert_{T_n}.
\end{align}
In the full 2d computation, the large gradient energy from the light scalar also makes this parameter much larger than that in the 1d approximation, as shown in Fig~\ref{fig:betaH compare fix tuning}.
One can see from the plot that the inverse duration parameter is independent of the light scalar mass in the 1d approximation, while in the full 2d computation it increases rapidly as $m_S$ goes down.
Here we scan over $m_S$ down to $0.01~\gev$.
For smaller $m_S$, the large change of the field value of $S$ in the bounce solution and the large $K_S$ leads to numerical difficulties, but we expect that $\beta/H$ increases further for smaller $m_S$.

The large $\beta/H$ for small $m_S$ can be understood in the following way. For a fixed temperature, the bounce action $S_3$ becomes larger for smaller $m_S$ because of the large change of the field value of $S\propto m_S^{-1}$. The action becomes smaller toward the nucleation temperature due to the cancellation between the positive contribution from the gradient term and the negative contribution from the potential term. The cancellation does not occur for the derivative of $S_3/T$ with respect to $T$, so $\beta/H$ becomes larger for smaller $m_S$.

\begin{figure}[t]
  \centering
  \includegraphics[width=0.6\linewidth]{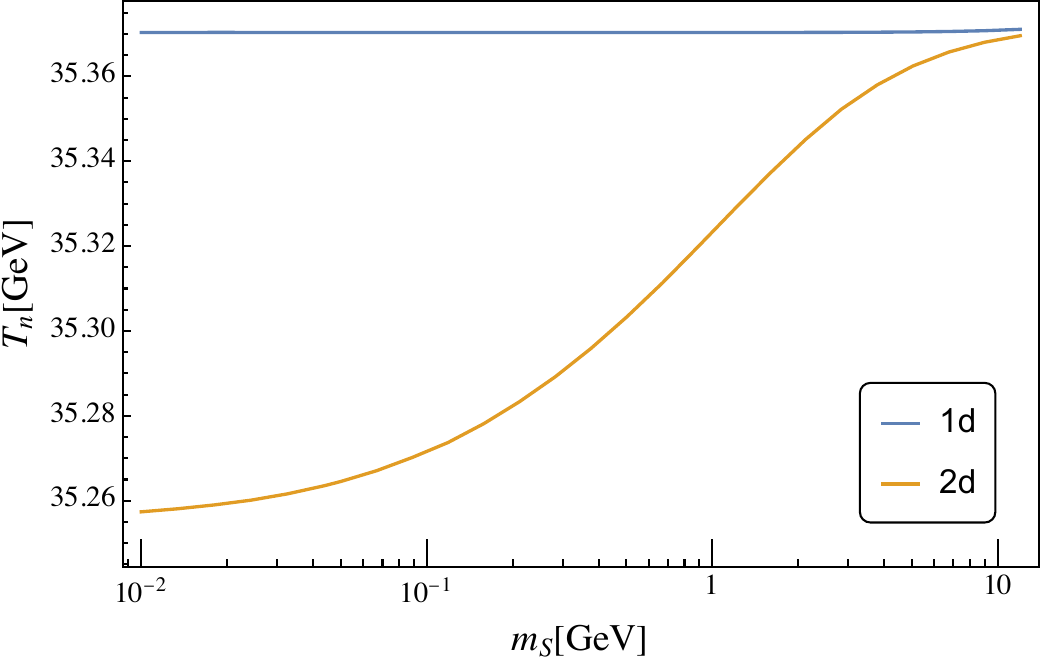}
  \caption{The nucleation temperature for fixed $T_c=35.42~\gev$.}
  \label{fig:Tn compare}
\end{figure}

The large $\beta/H$ has a profound impact.
Large $\beta/H$ corresponds to 
a rapid change of $S_3/T$ around the phase transition. Even if the bounce action is larger for smaller $m_S$, the nucleation is not significantly delayed. In Fig~\ref{fig:Tn compare}, we show how the nucleation is delayed for small $m_S$. The scan is along the line with fixed $T_c = 35.42~\gev$, which is the critical temperature for $10\%$ fine-tuning and $m_S \ll m_h$.
One can see that the nucleation is delayed very slightly at large $m_S$. The difference of the nucleation temperatures becomes larger as the mass decreases, then the mass dependence gradually becomes smaller as the mass decreases further.
The difference at $m_S\sim 10$ MeV is only around $0.1~\gev$. However, since the potential shape changes rapidly around $T_n$, the small change leads to a large increase of $v(T_n)$, which enhances the phase transition strength $v(T_n)/T_n$ by a factor of about $1.24$.
We expect that the ratio further increases for smaller $m_S$ because of smaller $T_n$ and the viable parameter region for SFOPT expands.
However, we find that the ratio $v(T)/T$ changes roughly linearly with the temperature, so the small change of the nucleation temperature below $m_S=10~\mathrm{MeV}$ should not lead to significant increase of $v(T_n)/T_n$. 
Since the parameters we scanned here are around the boundary of SFOPT, this ratio of $1.24$ helps us to figure out the boundary of SFOPT for $m_S\lesssim 10$ MeV, as discussed in the next subsection.

The short duration of the phase transition implies that the gravitational-wave signal produced from the bubble nucleation is suppressed. For example, we take the benchmark point $m_S = 5~\mathrm{MeV}$ and $\sin \theta = 1.2 \times 10^{-4}$, which is in the allowed region as we show in the next section.
Even if we take the bubble-wall velocity $v_b=1$ to maximize the GW signal, the peak of $\Omega_{\rm GW}h^2$ is at around $10^{-20}$ and located at the frequency around $10^2\rm Hz$, which is far below the sensitivity of the most sensitive proposed experiment, the Ultimate Decigo~\cite{Kudoh:2005as,Kuroyanagi:2014qaa}. 
Here we include the contribution from the sound-wave turbulence following~\cite{Caprini:2015zlo}, while the contribution from bubble collisions is subdominant in this parameter range and is neglected~\cite{Huber:2008hg}. 

\subsection{Nucleation vs critical}
\label{sec:Tn and Tc}

\begin{figure}[t]
  \centering
  \begin{minipage}[t]{0.49\linewidth}
    \hspace{0.5cm}
    \includegraphics[width=3.2in]{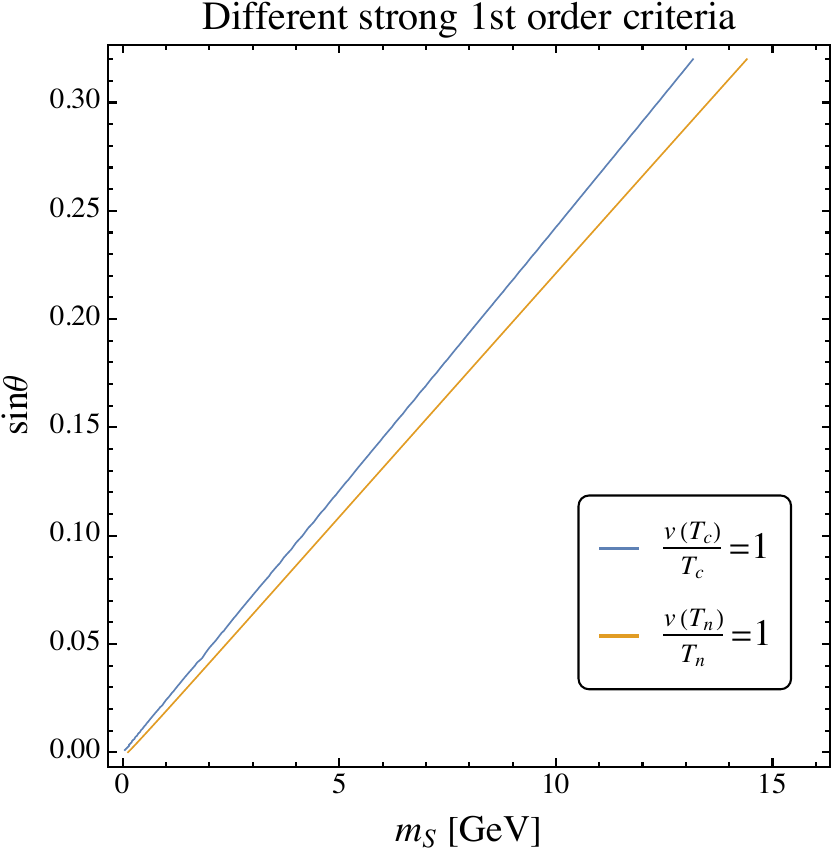}
  \end{minipage}
  \begin{minipage}[t]{0.49\linewidth}
    \hspace{0.5cm}
    \includegraphics[width=3.2in]{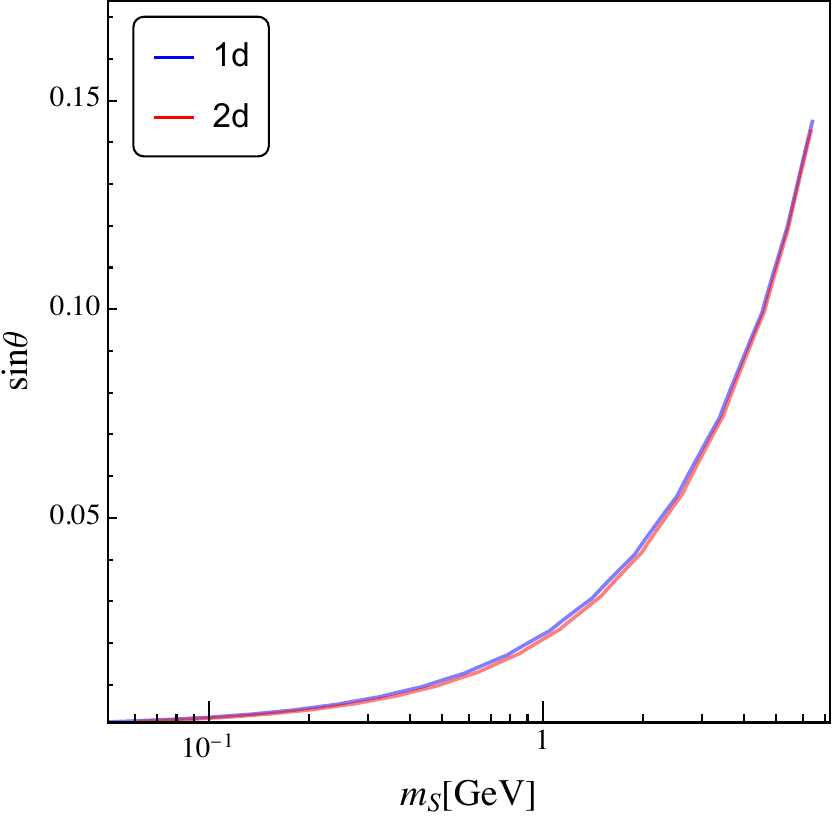}
  \end{minipage}
  \caption{Comparison between different SFOPT criteria. SFOPT is achieved above the lines. Left: Criteria imposed at the critical temperature $T_c$ and nucleation temperature $T_n$ obtained with the full 2d computation. Right: The condition $v(T_n)/T_n \geq 1$ imposed with the nucleation temperature from the 1d approximation and full 2d computation.}
  \label{fig:Tn vs Tc}
\end{figure}

A widely used condition for strong first order phase transition is given by $v_c/T_c\gtrsim 1$, which compares the vev of the Higgs with the temperature at the critical temperature $T_c$~\cite{Morrissey:2012db,Quiros:1999jp}. See~\cite{Moore:1998swa,DOnofrio:2014rug} for the computation of the sphaleron rate by lattice computation. This criterion is simple in a practical point of view.  However, the SFOPT is required in order to preserve the baryon number inside the bubble for EWBG, so what actually matters is the sphaleron rate in the broken phase at the nucleation temperature $T_n$.
A recent case study of the NMSSM shows that the condition at the nucleation temperature sometimes prefer a largely different parameter region than that at the critical temperature, especially for some very strong phase transitions where the barrier is difficult to conquer~\cite{Baum:2020vfl}.
In the left panel of Fig~\ref{fig:Tn vs Tc}, we compare these two different criteria ($T_c$ vs $T_n)$ for the full 2d computation. One can see that the condition at the nucleation temperature indeed increases the viable parameter region slightly.

As discussed before, the full 2d computation leads to a lower nucleation temperature and slightly larger phase transition strength. However, this difference does not lead to a large shift in the parameter region. In the right panel of Fig.~\ref{fig:Tn vs Tc}, we show the boundary of SFOPT condition $v(T_n)/T_n \geq 1$ for the 1d approximation and the full 2d computation. One can see that the shift of the viable parameter region is negligible.

In the following,
we adopt the condition $v(T_n)/T_n \geq 1$.
As we will see in the next section, the viable parameter region has $m_S< 10$  MeV,
and numerical issues appear as we stated earlier.
For such a small mass region, instead of actually solving the full 2D dynamics to obtain $v(T_n)/T_n$, we adopt the following approximation.
As can be seen from Fig.~\ref{fig:Tn compare}, the difference of the nucleation temperature  between the 1d and 2d computation is nearly independent of $m_S$ for $m_S < O(10)$ MeV with the fine-tuning around 10\%.
Also, for a fixed temperature $T$, $v(T)$ is the same for the 1D and 2D computation.
We thus scan the parameter space with the 1d approximation around the fine-tuning of $10\%$ to obtain $v(T_n)/T_n$, and use the following relation obtained in Sec.~\ref{sec:2dpt},
\begin{align}
  \label{eq:approximation}
  \left.\frac{v(T_n)}{T_n}\right\vert_{2d} \simeq 1.24 \left.\frac{v(T_n)}{T_n}\right\vert_{1d},
\end{align}
to infer $v(T_n)/T_n$ in the full 2D computation.

\section{Signals of the singlet}
\label{sec:signal}

In this section, we discuss observational and experimental signals of the singlet arising from the singlet-Higgs mixing. The large mass region $m_S > 5$ GeV is excluded by the LEP, the intermediate mass region $5 ~{\rm GeV}> m_S > O(10)$ MeV is excluded by rare B or K decay, and the low mass region $m_S < 3$ MeV is excluded by the observation of the CMB. The remaining parameter region, $3$ MeV $< m_S < O(10)$ MeV, can be probed by the future observation of rare K decay and the CMB.

\subsection{Scalar production at LEP}
\label{sec:scalar production}

The singlet-Higgs mixing induces an interaction $SZZ$, which leads to detectable collider signals. The extra scalar in the mass range $2~\gev \leq m_S \leq 15~\gev$ was searched at LEP~\cite{L3:1996ome} via $e^+ e^- \rightarrow Z \rightarrow S Z^*$ with the off-shell $Z^*$ and on-shell $S$ decaying into final state fermions, which constraints the mixing angle $\sin \theta$ to be smaller than $0.1$. From the left panel of Fig.~\ref{fig:combine}, we can see that this excludes all the parameter space with $m_S \geq 3.5~\gev$. Ref.~\cite{L3:1996ome} also considers the case where $S$ decays into invisible particles and put an even stronger bound on the mixing angle. Therefore, we cannot avoid the LEP bound by introducing a new invisible particle into which $S$ decays.   

\subsection{Rare Meson decay}

The mixing effect also leads to extra decay channels of mesons such as $B^0$, $B^+$, $K^0$, and $K^+$. See~\cite{Beacham:2019nyx} for a comprehensive review and~\cite{Goudzovski:2022vbt} for recent updates. Here we summarize the results relevant for our setup.

The $B$ meson can decay into $K$ and $S$, followed by $S$ decaying visibly into a muon pair. LHCb searches for 
$B^+ \rightarrow K^+ S (\mu^+ \mu^-)$~\cite{LHCb:2016awg}
and $B^0 \rightarrow K^0 S (\mu^+ \mu^-)$~\cite{LHCb:2015nkv} and constrains the mixing angle to be smaller than about $10^{-3}$ for $200~\mathrm{MeV} \leq m_S \leq 4~\gev$,
as shown in the left panel of Fig~\ref{fig:combine}.

A lighter mass range is probed via $K \rightarrow \pi S$ with the pion and $S$ further decay into visible or invisible final states~\cite{NA62:2020pwi,NA62:2021zjw}.
The limits are shown in both panels of Fig.~\ref{fig:combine}.
The computation of the branching ratio is based on the leading-order chiral perturbation~\cite{Leutwyler:1989xj}.
Since the Kaon mass is not much below the cut-off scale of the chiral perturbation theory, we consider that the limit has $O(1)$ uncertainty. The gap shown in Fig~\ref{fig:combine} can be probed in the next 10-15 years~\cite{Beacham:2019nyx} by NA62~\cite{NA62:2017rwk} and KLEVER~\cite{KLEVERProject:2019aks}.

\subsection{Effective number of neutrinos}
\begin{figure}[t]
  \centering
  \begin{minipage}[t]{0.495\linewidth}
    \hspace{0.5cm}
    \includegraphics[width=3.2in]{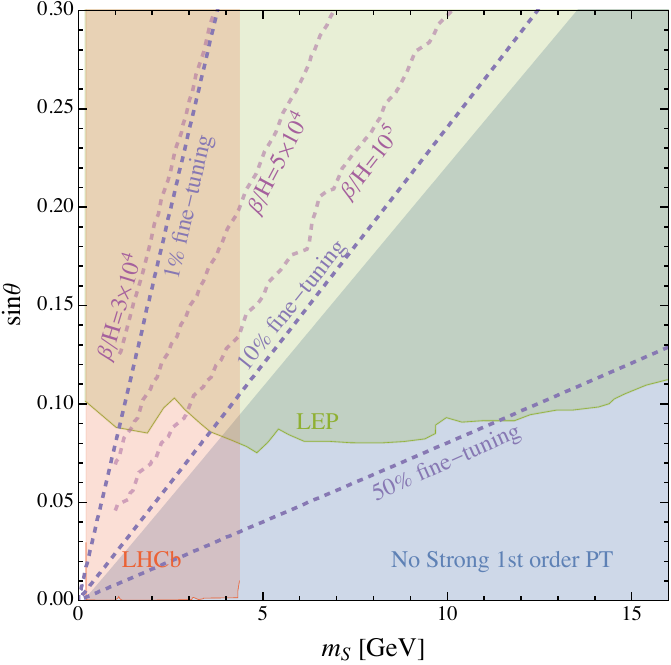}
  \end{minipage}
  \begin{minipage}[t]{0.495\linewidth}
    \hspace{0.5cm}
    \includegraphics[width=3.2in]{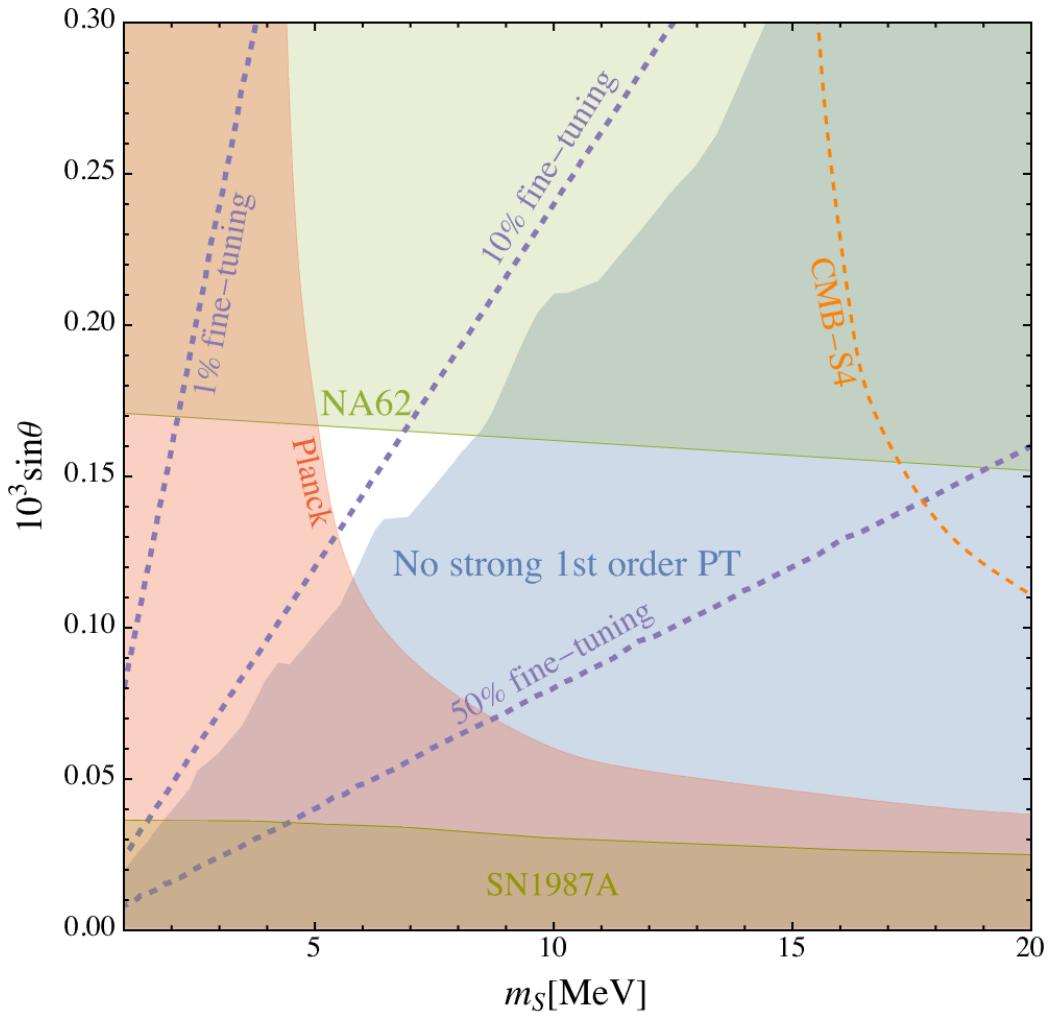}
    \end{minipage}
  \caption{Constraint on the parameter space. Left: Parameter space for GeV scale $m_S$, which is excluded by the LEP and LHCb. Right: Parameter space for MeV scale $m_S$, which can be probed by the observations of rare Kaon decay and the CMB. We also show the constraint from the cooling of SN1987A~\cite{Ishizuka:1989ts,Balaji:2022noj}.}
  \label{fig:combine}
\end{figure}

The singlet is in thermal equilibrium with the SM particles in the early universe. If $m_S \gg $ MeV, the energy density of the singlet becomes negligible by the time the neutrinos decouple from the bath, so the energy density of the relativistic species in the universe is not affected. If the mass is sufficiently close to the MeV scale, the singlet has a non-negligible amount of energy when the neutrinos decouple. The energy density is transferred only into the electron and photon, so the neutrinos become relatively cooler, leading to the effective number of neutrinos smaller than the SM prediction. The lower bound on the mass of an extra scalar that is in thermal equilibrium is derived in~\cite{Ibe:2021fed} using Planck 2018~\cite{Planck:2018vyg}, which is shown by a red-shaded region in the right panel of Fig.~\ref{fig:combine}.  CMB-S4 experiments~\cite{CMB-S4:2016ple} can probe the remaining viable parameter space (also derived in~\cite{Ibe:2021fed}), as is shown by a orange-dashed line.

\section{Electroweak baryogenesis by the singlet}
\label{sec:ewbg and edm}
In this section, we discuss a possibility of EWBG. We show that an enough amount of baryon asymmetry may be produced without generating too large an electron EDM. We also discuss implications of EWBG to the UV completion of the model.

\subsection{Local electroweak baryogenesis}
We introduce the following dimension-5 operator,
\begin{align}
    {\cal L} = \frac{\alpha_2}{8\pi} \frac{S}{M}W \tilde{W},
\end{align}
which can arise from the anomaly of the shift symmetry of $S$. Baryon asymmetry can be produced by the mechanism analogous to the local baryogenesis by the dimension-6 coupling between the Higgs field and the $W$ boson~\cite{Dine:1990fj}.
During the first order phase transition, the field value of $S$ changes and gives imbalance between the sphaleron and anti-sphaleron transition rates. We work in the thick-wall approximation, where the particles in the bath follow thermal distributions. The production rate of baryon asymmetry $n_B$ is
\begin{align}
    \dot{n}_B = 3 \Gamma_{\rm ws} \frac{\dot{S}}{M} T^2,~~\Gamma_{\rm ws} \simeq 20 \alpha_2^5 T.
\end{align}
Integrating this across the wall, we obtain
\begin{align}
    \frac{n_B}{s} \simeq \frac{3 \Gamma_{\rm ws} T^2 \Delta S}{M s } = \frac{3 \Gamma_{\rm ws} T^2 }{M s } \frac{A v(T)^2}{\mu_s^2} \simeq 10^{-10} \frac{10^8~{\rm GeV}}{M} \left(\frac{v(T_n)}{60~{\rm GeV}}\right)^2 \frac{10{\rm MeV}}{\mu_S},
\end{align}
where $\Delta S$ is the change of the field value of $S$ across the wall. We used $A^2 \simeq 2 \lambda \mu_S^2$ in the last equality. 
The observed baryon asymmetry may be explained for $M\sim 10^8$ GeV $\times (10~{\rm MeV}/\mu_S)$.

\begin{figure}[t]
  \centering
  \includegraphics[width=0.4\linewidth]{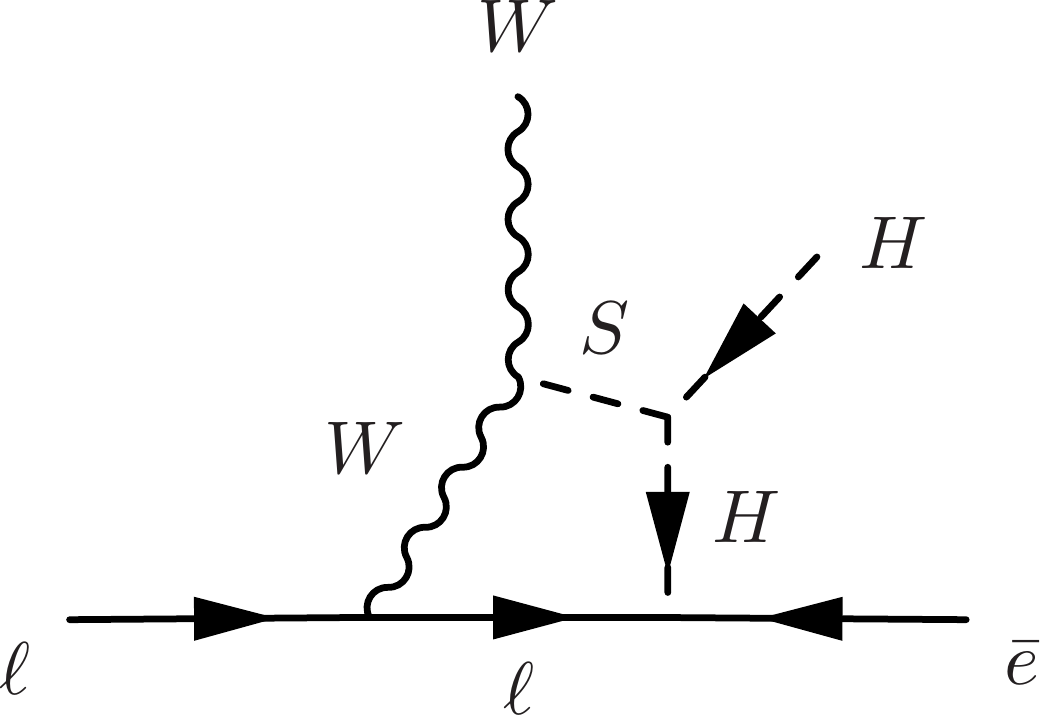}
  \caption{The diagram showing the quantum correction to the electron electric dipole moment.}
  \label{fig:EDM}
\end{figure}

The simultaneous existence of the dimension-5 coupling and $A S |H|^2$ breaks the CP symmetry, and the diagram in Fig.~\ref{fig:EDM} generates the following operator,
\begin{align}
    i F_{\mu \nu}\ell \sigma^{\mu \nu} \bar{e} H \times \frac{e y_e}{(16\pi^2)^2} \frac{A }{M v^2}.
\end{align}
The electron EDM arising from this operator is
\begin{align}
    d_e/e \simeq 10^{-36}~{\rm cm} \frac{10^8~{\rm GeV}}{M} \frac{\mu_S}{10 {\rm MeV}}.
\end{align}
For a fixed baryon asymmetry, i.e., fixed $M \mu_S$, the electron EDM is proportional to $\mu_S^2$. Therefore, the bound from the electron EDM, $d_e/e < 1.1\times 10^{-29}$ cm~\cite{ACME:2018yjb}, is satisfied in the viable parameter region $m_S \sim 10$ MeV.

\subsection{Implications to UV completion}
The successful baryogenesis has implications to UV completion of the theory. The scale $M$ is basically the decay constant of $S$, which is the UV completion scale of the $S$ sector. In the viable parameter region $m_S\sim 10$ MeV, $M \sim 10^{8}$ GeV is required. Note that $M \gg \Delta S$, so higher order terms in the potential of $S$, such as $S^3$, are negligible for the phase-transition dynamics.

The UV completion of the trilinear coupling would be given by
\begin{align}
\label{eq:OHH}
    c_{\cal O}{\cal O} H H^\dag, ~~A \sim c_{\cal O} \frac{\hvev{\cal O}}{M}.
\end{align}
where ${\cal O}$ is an operator whose phase direction is $S$.
Closing the Higgs loop,
with a cut-off scale $\Lambda_H$ of it,
the quantum correction generates%
\footnote{The quantum correction in Eq.~(\ref{eq:trilinear correction}) corresponds to that to the operator ${\cal O}^2$.}
\begin{align}
    \frac{c_{\cal O}}{16\pi^2}{\cal O} \Lambda_H^2,
\end{align}
which gives a correction to the mass of $S$,
\begin{align}
    \Delta \mu_S^2 \sim \frac{1}{16\pi^2} \frac{\Lambda_H^2}{M} \sim \frac{\lambda^{1/2}}{16\pi^2} \frac{\mu_S \Lambda_H^2}{M}. 
\end{align}
Requiring that this is smaller than $\mu_S^2$, we obtain an upper bound on $\Lambda_H$,
\begin{align}
\Lambda_H \lesssim 10^4~{\rm GeV}.
\end{align}
Here we used the relation between $M$ and $\mu_S$ for successful EWBG.
The cut-off scale may be much above the TeV scale, so LHC constraints on new particles are consistent with the naturalness of the lightness of $S$.
The cut-off might be as a result of supersymmetry breaking or compositeness of  the Higgs.

The cut-off may be instead introduced without supersymmetry or the compositeness of the Higgs, if the operator in Eq.~\eqref{eq:OHH} is suppressed for high momenta of the Higgs. 
Let us, for example, consider the following toy example,
\begin{align}
    y P \bar{L}_1 L_2  + \lambda_1 H \bar{N} L_1 + \lambda_2 H^\dag N \bar{L}_2 + m_1 \bar{L}_1 L_1 + m_2 \bar{L}_2 L_2 + m_N \bar{N}N,
\end{align}
where $P$ is a scalar whose angular direction is $S$, $N$ and $\bar{N}$ are singlet fermions and $L_{1,2}$ and $\bar{L}_{1,2}$ are doublet fermions. A one-loop correction generates
\begin{align}
    A \sim \frac{y \lambda_1 \lambda_2}{16\pi^2 }\frac{m_1 m_2 m_N}{\Lambda^2},~~\Lambda = {\rm max}(y \hvev{P},m_1,m_2,m_N),
\end{align}
and a two-loop correction generates
\begin{align}
    \Delta \mu_S^2 \sim \frac{y \lambda_1 \lambda_2}{(16\pi^2)^2}\frac{m_1m_2 m_N}{\hvev{P}}.
\end{align}
One can see that $\Lambda_H\sim \Lambda$, which is given by the fermion mass scale.
In this setup, the SM Higgs can be a fundamental field even up to the Planck scale. Low-scale supersymmetry is also not required.  The electroweak hierarchy may be explained by anthropic requirements~\cite{Agrawal:1997gf,Hall:2014dfa,DAmico:2019hih}. It will be also possible to embed the model into a theory where the electroweak scale is explained more naturally, such as composite Higgs models or supersymmetric theories. The little electroweak hierarchy problem in those theories will not disturb the lightness of the singlet.

\section{Metastability of the Higgs potential}
\label{sec:metastability}
The model successfully realizes SFOPT.
The smallness of the effective quartic coupling, $\lambda - A^2/(2\mu_S^2)$, is the key of this model. This coupling, however, receives quantum corrections from the top yukawa coupling and thus quickly becomes negative not far above the EW scale. As an example, for a benchmark point with $m_S = 5~\mathrm{MeV}$ and $\sin \theta = 1\times 10^{-4}$, the effective quartic coupling becomes negative at the energy scale around $400~\gev$.

The existence of the scale where the effective quartic becomes negative, however, does not necessarily imply a UV cut-off below the scale.
For example, the SM Higgs quartic coupling also becomes negative around the scale $10^{10}$ GeV~\cite{Buttazzo:2013uya}, but the tunneling rate from the EW vacuum to the unstable vacuum is much smaller than the Hubble constant at current time, so the SM does not have to be UV-completed below the instability scale.

Let us compute the tunneling rate in our setup. Tunneling proceeds via the nucleation of bubbles. The bubble nucleation rate per unit volume and time is
\begin{align}
  \label{eq:tunneling rate definition}
  \Gamma = \mathcal{A} e^{-S_E[\phi_i]},
\end{align}
where $S_E[\phi_i]$ is the Euclidean action of the bubble and the prefactor ${\cal A}$ is of mass dimension 4. We conservatively take ${\cal A}$ to be $(100~{\rm GeV})^4$, but since the light singlet is also involved, the prefactor may be smaller.

Any bubble configurations with an energy smaller than or equal to the initial homogeneous configuration can contribute to bubble nucleation, but the one with the smallest action dominantly contributes to the bubble nucleation process. 
The action is (locally) minimized by the $O(4)$ symmetric bounce solution~\cite{Coleman:1977py} if it exists. The bounce equation is
\begin{align}
  \label{eq:bounce equation}
- \frac{d^2\phi_i}{d r^2} - \frac{3}{r} \frac{d\phi_i}{dr} + \frac{\partial V}{\partial \phi_i} = 0.
\end{align}
The (at least locally) minimized Euclidean action is 
\begin{align}
  \label{eq:SE}
  S_E = 2 \pi^2\int r^3 dr \left( \sum_{i=h,S}\left (\frac{1}{2}\left(\frac{d \phi_i}{dr}\right)^2\right ) + V(\phi(r)) - V(\phi_F) \right),
\end{align}
where $\phi_F$ is the field value at the false vacuum.

\begin{figure}[t]
  \centering
  \includegraphics[width=0.5\linewidth]{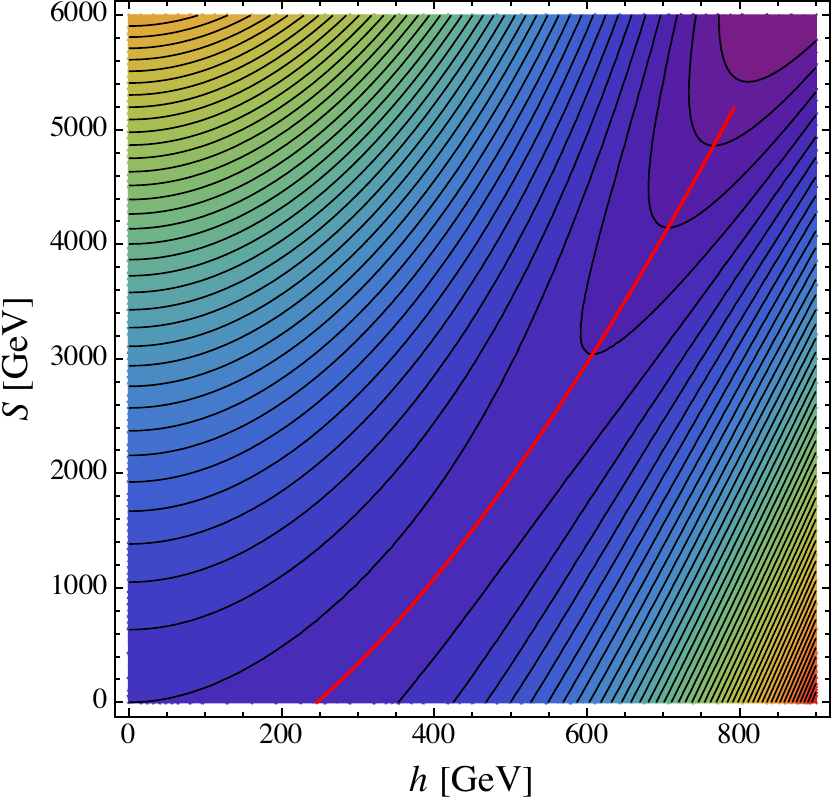}
  \caption{Tunneling trajectory from the EW vacuum to infinity for $m_S = 5~\gev$ and $\sin \theta = 0.17$. The red curve shows the tunneling trajectory. Black contours show the value of the effective potential at $T=0$. The trajectory is along the ``valley'' of the potential where $\partial V/\partial S = 0$.}
  \label{fig:path}
\end{figure}

We obtain the bounce solution of our setup via \verb|SimpleBounce|~\cite{Sato:2019wpo}. For the purpose of illustration, here we show the tunneling trajectory of a benchmark point $m_S = 5~\gev$, $\sin \theta = 0.17$ in Fig~\ref{fig:path}. The trajectory starts at the EW scale vev $(246~\gev, 0~\gev)$ and almost follows the valley defined by $\partial V/\partial S = 0$. A potential bump is conquered during this process, and the tunneling action is of order $10^{6}$, which implies a stable vacuum at the EW scale. We computed the bounce action in the parameter space with $m_S \in [2~\gev, 15~\gev]$ and $\sin \theta \in [0.1, 0.3]$, and found that the bounce action is at least order $10^{4}$. The large action is due to the large displacement of the field value of $S$ in the bounce solution, which leads to a large gradient energy.

\begin{figure}[t]
  \centering
  \includegraphics[width=0.6\linewidth]{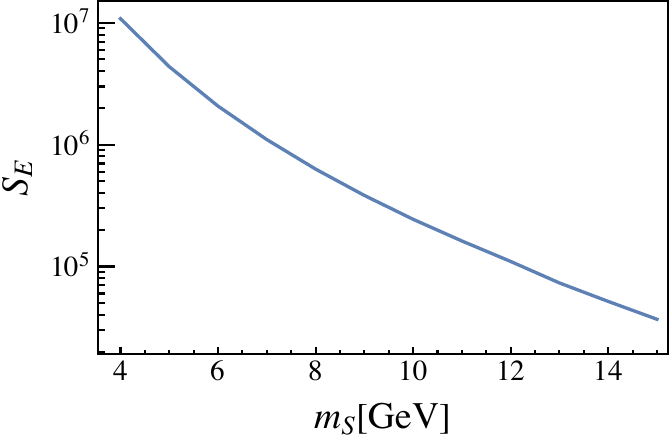}
  \caption{Tunneling action as a function of $m_S$ for the fine-tuning of $5\%$.}
  \label{fig:tunnling action}
\end{figure}

For the viable parameter region where $m_S$ is below 10 MeV, numerical issues again appear due to the large kinetic energy and large mass hierarchies. However, the tunneling action there is expected to be much larger than that in $m_S \gg 10$ MeV. In Fig~\ref{fig:tunnling action}, we show the tunneling action as a function of $m_S$ for fixed fine-tuning of $5\%$.
The tunneling action blows up as $m_S$ decreases in proportion to $m_S^{-4}$. This behavior can be understood by estimating the tunneling action by a dimensional analysis on the path with $\partial V/\partial S = 0$ while neglecting the kinetic energy from $H$.

The bounce action, however, not necessarily gives the dominant contribution to the nucleation process.
Indeed, the bounce solution can only give a local minimum value of the action instead of a global one.
For example, one can take a path along $S=0$, for which the bubble is the same as that in the SM, which has a much smaller action~\cite{Fubini:1976jm,Lipatov:1976ny},
\begin{align}
    S_E \simeq \frac{8\pi^2}{-3 \lambda (h_b)},
\end{align}
where $h_b$ is the Higgs field value inside the bubble large enough so that $\lambda(h_b) <0$.  Although the action is much smaller than the bounce action shown above, the tunneling rate by this bubble is still small and the EW vacuum is stable enough.%
\footnote{
Except for $h_b$ that maximizes $|\lambda(h_b)|$, the bounce solution does not exist.
Still, the pseudo-bounce action method~\cite{Espinosa:2019hbm} gives the minimal action for a given $h_b$.
}

One may wonder if taking a path with non-zero $S$ can further reduce the action and destabilize the EW vacuum. We argue that that will not the case. In order for the action to be significantly reduced by a non-zero singlet field value, the trilinear term $A S |H|^2$ must be comparable to the quartic term $\lambda |H|^4$. However, because of the smallness of $A$, this requires $S \gg H$, leading to a larger action from the kinetic term of $S$.

\section{Conclusions and Discussion}
\label{sec:discussion}

In this paper, we investigated an extension of the SM by a singlet scalar field. The singlet has a softly broken shift symmetry and hence is naturally light. The leading potential terms are a mass term and a trlinear coupling with the SM Higgs. The EWPT can be of strong first order.

The model has only two free parameters. We investigated the phase transition dynamics and identified the parameter space exhibiting a strong first order phase transition.
The parameter space roughly coincides with what is obtained by simply requiring that the vev of the Higgs at the the critical temperature is above the temperature. 
For smaller singlet masses, the duration of the phase transition becomes shorter and the magnitude of the gravitational waves produced by the first order phase transition is suppressed.

We also checked the constraints from collider searches, rate decay of mesons, and the CMB. The viable parameter space has a singlet mass around 10 MeV and can be probed by future observations of Kaon decay and the CMB. Unfortunately, for such a small mass, the duration of the phase transition is so short that the gravitational waves produced by the phase transition is too weak to be detected.

The model has an apparent instability at the Higgs field value below $1$ TeV. However, the tunneling into the true vacuum involves a large change of the field value of the singlet. As a result, the tunneling rate is suppressed and the electroweak vacuum is stable enough. No new physics that couples to the Higgs below the TeV scale is required.

If the singlet couples to the weak gauge boson, the baryon asymmetry of the universe can be explained by local electroweak baryogenesis. The predicted electron EDM is too small to be detected by near-future experiments. In this setup, to keep the singlet light, there should be new physics that cuts-off the quantum correction from the SM Higgs below $O(10)$ TeV.

Interestingly, the setup cannot be further probed by any of the conventional probes of EWBG; collider experiments, EDM measurements, nor gravitational-wave observatories. Instead, as mentioned above, the model can be probed by less-conventional probes; rare Kaon decay and the CMB.

\section*{Acknowledgement}

We thank Djuna Croon, Teppei Kitahara, Zhen Liu, Kohsaku Tobioka, and Yikun Wang for useful discussion. IRW is supported by the DOE grant DE-SC0010008.

\appendix

\section{Parametrization}
\label{sec:parametrization}

In this appendix, we show how the parameters $\{\lambda, \mu_H, \mu_S, A\}$ are connected with another set of parameters $\{m_h, m_S, v, \sin \theta\}$ at the tree and 1-loop level.
The Higgs mass and the EW vev are fixed as $m_h = 125~\gev$ and the EW vev $v = 174~\gev$.
The remaining two parameters, $m_S$ and $\sin \theta$, are free parameters of the model.

At the tree-level, the mass matrix in Eq.~\eqref{eq:mass matrix} is computed from the tree-level potential. Matching the masses and mixing angle and using the condition that the tree-level vev is $h=\sqrt{2}v$ and $S=0$, we obtain
\begin{align}
  \label{eq:parametrization}
  \mu_H^2 &=  \frac{1}{2} \left( m_h^2\cos^2\theta + m_S^2\sin^2\theta \right),~~
  \mu_S^2 = m_S^2 \cos^2 \theta + m_h^2 \sin^2 \theta, \nonumber \\
  A &= \frac{\left( m_h^2 - m_S^2 \right)\sin 2\theta}{2 \sqrt{2}v},~~
  \lambda = \frac{m_h^2 \cos^2\theta + m_S^2\sin^2\theta }{4v^2}.
\end{align}
We use this relation for the scanning in the parameter space.

At 1-loop level, the Coleman-Weinberg correction to the effective potential is computed in Eq.~\eqref{eq:CW potential}. For the zero temperature effective potential $V_{1} = V_0 + V_{\rm CW}$, the scalar masses are computed from the mass matrix
\begin{align}
  \label{eq:mass matrix}
  M_{\rm scalar} =
  \begin{pmatrix}
    \frac{\partial^2}{\partial h^2}V_{1} & \frac{\partial^2}{\partial h \partial S}V_{1}\\
    \frac{\partial^2}{\partial h \partial S} V_{1} & \frac{\partial^2}{\partial S^2}V_{1}
  \end{pmatrix}\bigg\rvert_{h=\sqrt{2}v, S=0}.
\end{align}
The eigenvalues of the mass matrix correspond to the physical masses $m_h$ and $m_S$. The mixing angle $\sin \theta$ is given by the relative angle between the eigenvectors and the gauge-basis vectors.
So far we have three relations.
The fourth relation is
\begin{align}
  \label{eq:boundary condition}
  \frac{\partial}{\partial h} \left( V_0 + V_{\rm CW} \right)|_{h=\sqrt{2}v,S=0} = 0.
\end{align}
With these four relations, we can translate 
$\{\lambda, \mu_H, \mu_S, A\}$ into $\{m_h, m_S, v, \sin \theta\}$.
This relation is used for the discussion on the vacuum stability in Sec.~\ref{sec:metastability}.

\small{\bibliography{SFOPT_light_scalar}}

\end{document}